\documentclass[preprint,showpacs,preprintnumber,amsmath,amssymb,aps]{revtex4}
\usepackage{graphicx}

\begin{document}

\title{Electron-acoustic-phonon scattering and electron relaxation in two-coupled quantum rings}

\author{G. Piacente} \email{gpiacente@ursa.ifsc.usp.br}
\author{G.~Q. Hai}

\affiliation{Instituto de F\'{\i}sica de S\~ao Carlos, Universidade
de S\~ao Paulo, 13560-970 S\~ao Carlos, S\~ao Paulo, Brazil}

\begin{abstract}
Electron relaxation, induced by acoustic phonons, is studied for
 coupled quantum rings in the presence of external fields, both
electric and magnetic. We address the problem of a single electron
in vertically coupled GaAs quantum rings. Electron-phonon
interaction is accounted for both deformation potential and
piezoelectric field coupling mechanisms. Depending on the external
fields, the ring radii and the separation between the rings, we show
that the two different couplings have different weights and
importance. Significant oscillations are found in the scattering
rates from electron excited states to the ground state, as a
function of either the geometry of the system or the external
fields.

\end{abstract}


\maketitle

\section{Introduction}

Ring geometries have long fascinated physicists for their peculiar
features. Recent successful advances in nano-fabrication technology
have allowed the realization of self-assembled heterostructured
semiconductor rings of nanometer size
\cite{garcia,lorke,mailly,mano}. These quantum objects, which are
nearly disorder free and contain only a few interacting electrons,
are attracting a great deal of interest in theoretical and
experimental research. Quantum rings (QRs) represent an alternative
to quantum dots (QDs) as zero-dimensional structures for practical
use as nanodevices. Their not simply-connected geometry provides
them with distinctive electronic structures, magnetic field
responses and transport properties, which could be of great
practical interest.

In the last years much attention has been dedicated to the
properties of QDs and coupled quantum dots (CQDs) and their physics
is by now well understood. On the other hand, only recently their QR
counterparts have started being addressed. A number of experimental
and theoretical works have studied laterally-coupled
\cite{Planelles}, concentrically-coupled
\cite{ManoPRB,SzafranPRB,PlanellesEPJB}, vertically-coupled
\cite{GranadosAPL,JClimentePRB,MaletPRB,Castelano} and
stacked\cite{hun} nanoscopic QRs.

Since the proposal of a qubit based on the electronic states of QDs
and CQDs \cite{bayerSC,burkard}, much work has been done in order to
understand the carrier-relaxation processes in QDs and CQDs, because
a long coherence time is required. The relaxation via phonon
emission has received widespread attention. This intrinsic mechanism
affects fundamental properties of semiconductor nanostructures and
may be the only non-radiative process controlling electron energy
losses. Among various aspects of phonon-assisted relaxation, the
phonon bottleneck effect, predicted in Ref. \cite{BastardPRB}, was
carefully studied. It is the basis of predicted large reductions of
electron relaxation rates in QDs when compared to two-dimensional (2D) or one-dimensional (1D)
heterostructures. Due to the discrete electron energy spectrum, the
interaction with phonons occurs only when the interlevel separation
closes to the energy of longitudinal optical (LO) phonons or it is
smaller than the bandwidth of acoustic phonons (a few meV). In
lateral or vertical QDs, where the energy level separation is small
as compared to the optical phonon energies, the electron-acoustic
phonon interaction is dominant. In coupled quantum rings (CQRs) the
energy separation is of the order of few meV \cite{MaletPRB}, so the
relaxation through emission of acoustic phonons is expected to be
the most efficient relaxation mechanism.

Electron relaxation in QDs and CQDs due to acoustic phonon
scattering has been studied extensively over the past few years
\cite{FujisawaNAT,OrtnerPRB,ClimentePRB}. Acoustic phonon induced relaxation has not yet been
studied in vertically CQR geometries up to now. In a recent work
\cite{piacente} we investigated the electron relaxation induced by
acoustic phonons in a single electron QR in the presence of an
external magnetic field. We considered both the deformation
potential (DP) and piezoelectric (PZ) acoustic phonon scattering.
Motivated by the recent experimental realization of complexes
consisting of stacked layers of InGaAs/GaAs QRs
\cite{granadosapl,suareznano}, here, we face the problem for two
vertically coupled quantum rings in the presence of external
electric and magnetic fields. Moreover, understanding and
controlling the electron relaxation rates is crucial in view of
future CQR applications,  such as efficient lasers and
opto-electronic devices, and in order to achieve high resolution
optical spectroscopy, where phonon scattering rates smaller than
photon emission and absorption rates are required. Therefore, the
knowledge of the physics of electron coupling to acoustic phonon in
QRs and CQRs is of great theoretical and practical interest.

The present paper is organized as follows. In Sec. \ref{sec2} we
summarize the model of CQR we adopt and describe the different
electron-phonon coupling mechanisms and the numerical methods we use
to calculate the relaxation times. Section \ref{sec3} contains the
numerical results as well as discussions and comments. Finally, we
conclude in Sec. \ref{sec4}.

\section{Model and methods}\label{sec2}

As a prototypical system we consider a GaAs/AlGaAs coupled quantum
ring structure, formed by two identical rings coupled along the
growth direction, i.e. the $z$-direction. The system as a whole has
rotational symmetry along $z$. We model the two coupled rings with a
displaced parabolic confinement in the $x-y$ plane and a symmetric
double-well in the $z$-direction: $V({\bf r})=
\frac{1}{2}m^*\omega_0^2({\bf r}_\parallel-r_0)^2 + V_z$, where
$r_0$ is the ring radius and $V_z(z)= V_l$ if $W_b/2 \leq |z| \leq
(W_b/2 + W_z)$ and $V_z(z)= V_h$ otherwise. Here, $W_z$ is the
thickness of the GaAs layers and $W_b$ is the thickness of the
inter-ring layer; $V_0 \equiv V_h-V_l$ is the conduction band-offset
of GaAs/AlGaAs. Furthermore, we consider either a magnetic ($B$) or
an electric ($E_z$) field applied in the $z$-direction. Since the
confinement in the vertical direction is usually much stronger than
the lateral one, the in-plane motion and the vertical one can be
treated as decoupled, therefore the eigenfunctions are given in the
separable form $\psi_{nmg}({\bf r})= \phi_{nm}(x,y) \chi_{g}(z)$,
with the $n=0,1,2\dots$, $m=0,\pm 1,\pm 2, \dots$ (the angular
quantum number) and $g=0,1,2,\dots$ The functions $\phi_{nm}(x,y$)
are linear combinations of the Fock-Darwin states for the QD (for
details see Refs. \cite{piacente,simoninPRB}), while the functions
$\chi_{g}(z)$ are the eigenfunctions of the double-well problem. In
what follows we will consider only the $g=0$ (bonding) and $g=1$
(antibonding) states in the vertical direction. Bonding
(antibonding) states have even (odd) parity with respect to the
reflection about the $z=0$ plane.

The electron-phonon scattering rate at zero temperature, where
phonon absorption and multi-phonon processes are negligible, can be
obtained by the Fermi golden rule, as long as the energy difference
between two electron levels is much smaller than the LO-phonon
energy \cite{HaiAPL}. This is the case we study here with $\hbar
\omega_0$ being a few meVs. The scattering rate between the initial
state $\psi_i$ and the final state $\psi_f$ is given by

\begin{equation}
\tau^{-1}_{i \rightarrow f}=\frac{2\pi}{\hbar}\,\sum_{\lambda,
\mathbf{q}} | M_{\lambda}(\mathbf{q})|^2\, |\langle \psi_f|\,e^{-i
\mathbf{q} \cdot \mathbf{r}}\,|\psi_i \rangle|^2\, \delta(|E_f-E_i|
- E_q), \label{eq1}
\end{equation}

\noindent where $M_{\lambda}(\mathbf{q})$ is the scattering matrix
element corresponding to different electron scattering mechanisms
$\lambda$, $\mathbf{q}$ the phonon wave number, $E_f$ and $E_i$ the
final and initial electron state energies, respectively, and $E_q$
represents the phonon energy. It is evident from Eq. (\ref{eq1}) that
the relaxation is mediated by phonons whose energy matches that of
the transition between the initial and final electron states.

In a polar semiconductor such as GaAs, electrons couple to all types of
phonons, i.e. electrons couple to longitudinal acoustic (LA) phonons
through a deformation potential and to longitudinal and transverse
acoustic (TA) phonons through piezoelectric interactions
\cite{Mahan}. The total scattering matrix element is the sum of all
the different contributions.

The electron-LA phonon scattering due to the deformation potential
has the form:

\begin{equation}
| M_{\mbox{\tiny LA}}^{\mbox{\tiny DP}}({\mathbf{q}}) |^2 =
\frac{\hbar D^2}{2 \rho v_l \Gamma} |{\mathbf{q}}|,
\end{equation}

\noindent where $D$, $\rho$, $\Gamma$, and $ v_l$ are the crystal
acoustic deformation potential constant, the density, the volume and
the longitudinal sound velocity, respectively. For GaAs (zinc-blende
structure) the only nonvanishing independent piezoelectric constant
is $h_{14}$ and the coupling function due to the piezoelectric interaction is given by:

\begin{equation}
| M_{\mbox{\tiny LA}}^{\mbox{\tiny PZ}}({\bf q}) |^2=\frac{32
\pi^2\,\hbar e^2 h_{14}^2}{\epsilon_0^2 \rho v_l \Gamma}\frac{(3 q_x
q_y q_z)^2}{|{\bf q}|^7},
\end{equation}

\noindent for the electron - LA phonon scattering and

\begin{equation}
|M_{\mbox{\tiny TA}}^{\mbox{\tiny PZ}}({\bf q }) |^2 = \frac{32
\pi^2 \hbar e^2 h_{14}^2}{\epsilon_0^2 \rho v_t \Gamma}
\bigg|\frac{q_x^2 q_y^2 + q_y^2 q_z^2 + q_z^2 q_x^2}{|{\bf q }|^5} -
\frac{(3 q_x q_y q_z)^2}{|{\bf q}|^7}\bigg|
\end{equation}

\noindent for the electron - TA-phonon scattering. The transversal
sound velocity is written as $v_t$. We used linear approximations
$\omega_{q}^{\mbox{\tiny LA}}=v_l q$ and $\omega_{q}^{\mbox{\tiny
TA}}=v_t q$ for the LA and TA phonon dispersion, respectively. In
our calculation we used GaAs/Al$_{0.3}$Ga$_{0.7}$As material
parameters: electron effective mass $m^*=0.067$, band-offset
$V_0=240$ meV, $\rho=5300$ kg/m$^3$, $D=8.6$ eV, $\epsilon=12.9$,
and $h_{14}=1.4 \times 10^9$ V/m. For the sound speeds we used the
values $v_l=3.7 \times 10^3$ m/s and $v_t=3.2 \times 10^3$ m/s
\cite{blackmore,piacente} .

In this work we consider mostly relaxation rates from the first
excited to the ground electron state. Since many applications rely
on the creation of a two-level system, the relaxation of an electron
from the first excited state is often the most relevant transition
and, furthermore, it can be monitored, e.g., by means of
pump-and-probe techniques \cite{FujisawaNAT}. The behavior of
transition rates from higher excited states is qualitatively
similar, except for the presence of a larger number of decay
channels which smears the features of direct scattering between two
selected states, as discussed e.g. in Ref. \cite{piacente}.

\section{Results and discussion}\label{sec3}

\begin{figure}[h]
\begin{center}
\includegraphics [width=7.5cm]{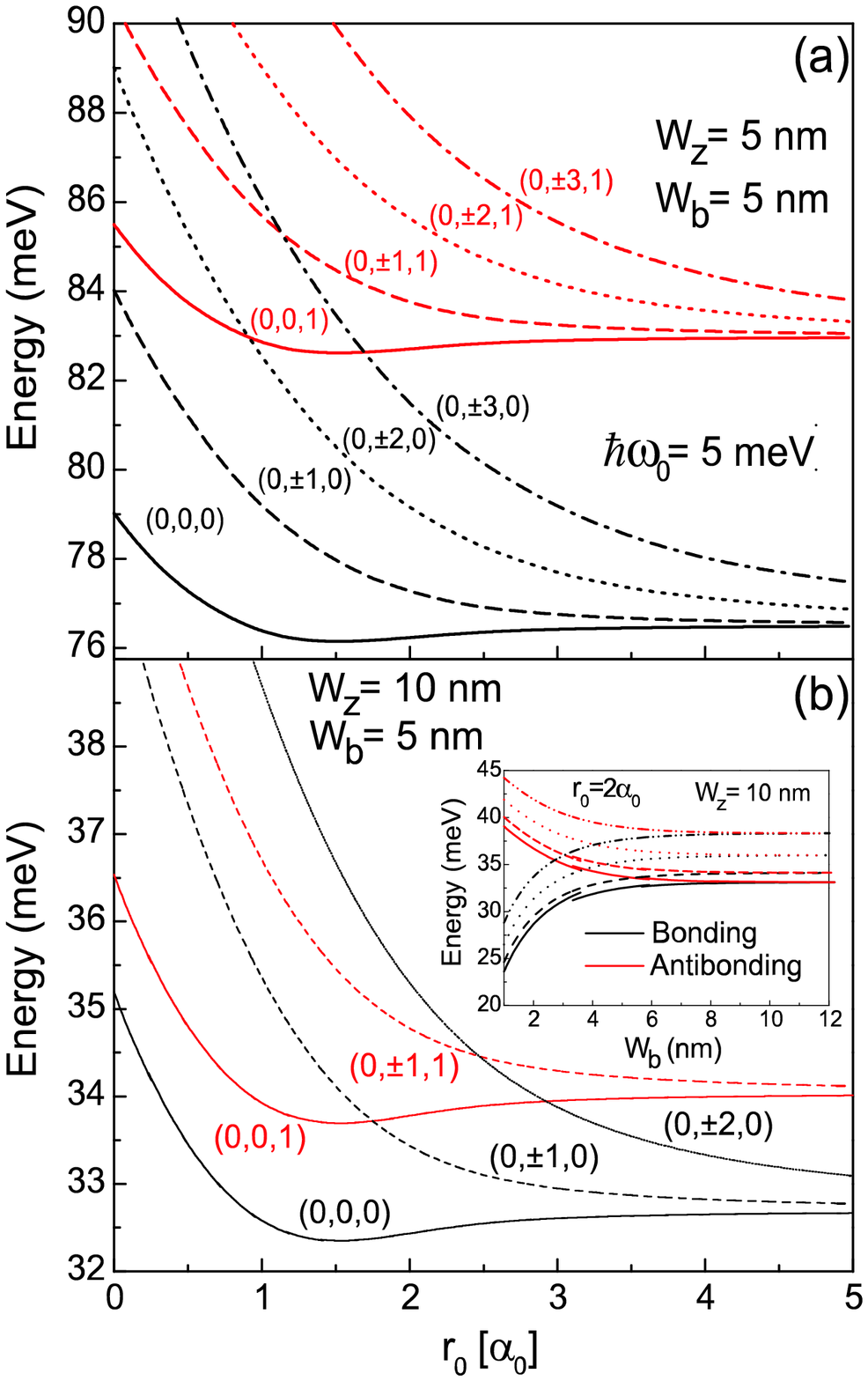}
\protect\caption{(Color online) The energy spectrum for $B=0$ as a
function of the ring radius at fixed lateral confinement
$\hbar\omega_0=5\,$meV: (a) for $W_b=5\,$nm and $W_z=5\,$nm; (b)
$W_b=5\,$nm and $W_z=10\,$nm. In (b) the inset shows the lowest
bonding and antibonding levels as a function of the ring separation
$W_b$ for $r_0=2\alpha_0$.} \label{fig1}
\end{center}
\end{figure}

In order to identify the ground and the first excited electron
states, we plot in Fig. \ref{fig1} the energy levels as a function
of the ring radius for different vertical confinements in the
absence of external fields.  We measure the ring radius in unit
$\alpha_0 = \sqrt{\hbar/m^{\ast}\omega_{0}}$, which for a lateral
confinement $\hbar \omega_0=5\,$meV results in a unity length of
about 15 nm.

It is interesting to notice the differences between the single QR
and CQR cases. The CQR spectrum is much richer than the single QR
one due to the presence of antibonding levels which crosses with
the bonding ones, a feature which is not present in single QRs. In
CQRs even in the absence of magnetic field the ring radius can alter
the quantum numbers of excited states. In particular, for $W_z=10\,$
nm the first excited state is a piecewise function of the ring
radius, made by the $(0,0,1)$ and $(0,\pm 1,0)$ levels. It is also
worth noting that a stronger vertical confinement results in a
shift to higher energy values for fixed values of the barrier
thickness $W_b$ and ring radius. In the inset of Fig. \ref{fig1}(b),
we show the lowest energy levels as a function of $W_b$ for a fixed
value of the ring radius. One can observe that the energy difference
between bonding and antibonding states is a decreasing function
with increasing ring separation.

\begin{figure}[h]
\begin{center}
\includegraphics [width=7.5cm]{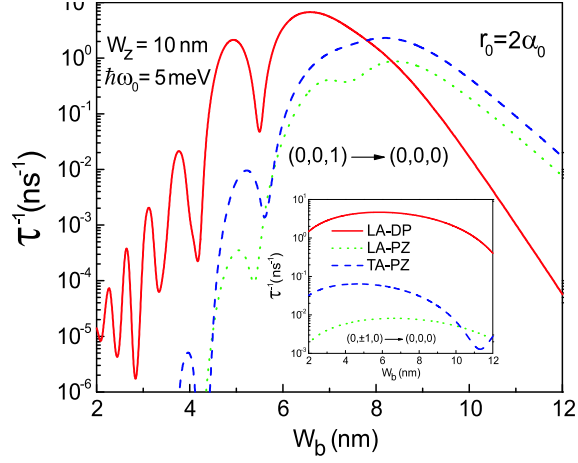}
\protect\caption{(Color online) The LA-DP, LA-PZ and TA-PZ
scattering rates for the $(0,0,1)\rightarrow(0,0,0)$ transition as a
function of the barrier thickness $W_b$. The inset shows the
scattering rates for the $(0,\pm 1,0)\rightarrow(0,0,0)$
transition.}\label{fig2}
\end{center}
\end{figure}

\begin{figure}[h]
\begin{center}
\includegraphics [width=7.5cm]{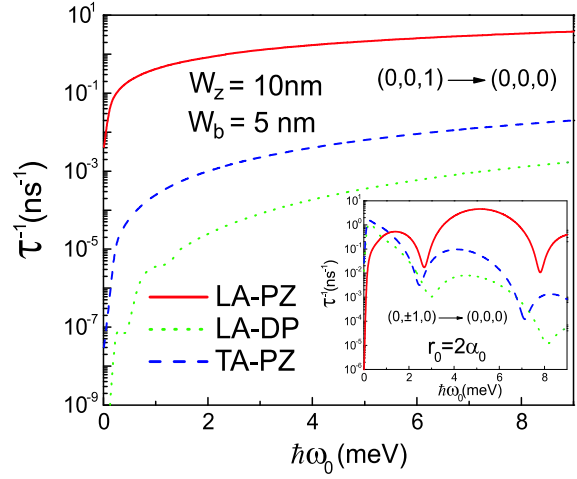}
\protect\caption{(Color online) The LA-DP, LA-PZ and TA-PZ
scattering rates for the $(0,0,1)\rightarrow(0,0,0)$ transition as a
function of the lateral confinement $\hbar \omega_0$. The inset
shows the scattering rates for the $(0,\pm 1,0)\rightarrow(0,0,0)$
transition.} \label{fig3}
\end{center}
\end{figure}

Electron-acoustic phonon relaxation in CQRs is to a large extent
determined by the interplay of lateral and vertical confinement
strengths relatively to the single rings, inter-ring thickness, and
emitted phonon wavelength.  Figure \ref{fig2} shows the scattering
rates for the $(0,0,1) \rightarrow (0,0,0)$ and $(0,\pm 1,1)
\rightarrow (0,0,0)$ (the inset) transitions as a function of $W_b$
in the CQRs with $\hbar \omega_0=5 \,$meV and $W_z=10$ nm, for the
different scattering mechanisms in the absence of external fields.
In particular, we considered values of the $W_b$  from 2 to 12 nm,
accordingly to recent experimental realizations of stacked layers of
self-assembled QRs, where the inter-ring thickness were 1.5, 3, 4.5,
6, 10 and 14 nm \cite{granadosapl,suareznano}. The scattering rate
oscillates strongly for the $(0,0,1) \rightarrow (0,0,0)$ transition
as a function of $W_b$, on the other hand such oscillations are not
found in the $(0,\pm 1,0) \rightarrow (0,0,0)$ transition. The reason for these
oscillations is that the electron wave function along the
$z$-direction can be in-phase or in anti-phase with the phonon wave,
in particular, there are maxima when the electron wave function is
in-phase with the phonon wave and minima when the electron wave
function is in anti-phase with the phonon wave. The tunneling energy
in CQRs depends on the barrier thickness $W_b$ between the two
rings: tunneling between the rings affects strongly the
electron-phonon interaction because the electron wave function
spreads in the two rings leading to different phonon wavelengths
matching the electron-phonon interference. For the sake of clarity,
the energy, and thus the wavelength of the emitted phonon, of the
$(0,0,1)\rightarrow(0,0,0)$ transition from an antibonding to a
bonding state is exclusively related to the inter-ring thickness,
while the energy of the $(0,\pm1,0)\rightarrow(0,0,0)$ transition
from a bonding to another bonding state is related to the lateral
confinement only. Therefore, if we fix the lateral confinement and
change the inter-ring thickness, we get strong oscillations in the
scattering rate for the $(0,0,1)\rightarrow(0,0,0)$ transition. On
the other hand, if we fix the inter-ring thickness and change only
the lateral confinement, we get strong oscillations in the
scattering rate for the $(0,\pm1,0)\rightarrow(0,0,0)$ transition
(see Fig. \ref{fig3}).

From Fig. \ref{fig2} one can see that for small ring separations in
the $(0,0,1) \rightarrow (0,0,0)$ transition the LA-DP scattering is
much more efficient than the LA-PZ and TA-PZ ones, while for larger
values of the barrier thickness the LA-PZ and TA-PZ couplings start
to dominate over the LA-DP one. However, for the $(0,\pm
1,0) \rightarrow (0,0,0)$ transition the LA-DP scattering is always larger than
the other ones. Furthermore, the TA-PZ scattering is generally
larger than the LA-PZ one, except for large $W_b$.

In what follows we concentrate on geometries characterized by
$W_z=10\,$nm and $W_b=5\,$nm. Figure \ref{fig3} illustrates the
scattering rate for a CQR system as a function of the lateral
confinement energy $\hbar \omega_0$. For most lateral confinements
and for $r_0=2 \alpha_0$, DP coupling gives the largest contribution
and the PZ coupling is negligible. However, for very weak
confinements in the $(0,\pm 1,0) \rightarrow (0,0,0)$ transition
(see the inset), the PZ coupling prevails. This is a result in
agreement with recent findings in CQDs \cite{ClimentePRB}. As
already mentioned, strong oscillations in the scattering rate are
present in the $(0,\pm 1,0) \rightarrow (0,0,0)$ transition, but not
in the $(0,0,1) \rightarrow (0,0,0)$ transition, just the opposite
with respect to the case of the dependence of barrier thickness. In
the presence of strong oscillations the scattering rate is
suppressed by orders of magnitude. This phenomenon, already observed
in QDs and CQDs \cite{ClimentePRB} and in QRs \cite{piacente}, has
been proposed as a possible way to preserve the quantum coherence in view
of the implementation of quantum computing devices based on
semiconductor nanostructures \cite{zanardiprl}.

Although the shape of PZ scattering rate is different from the DP
one as a function of $\hbar \omega_0$, the limiting behavior is
similar: they tend to zero at very weak confinement potentials
because of the vanishing phonon density. Moreover, we observe again
that the PZ contribution coming from the TA phonons is larger than
that of the LA phonons. This holds for almost all the calculations
throughout this paper.

\begin{figure}[h]
\begin{center}
\includegraphics [width=7.5cm]{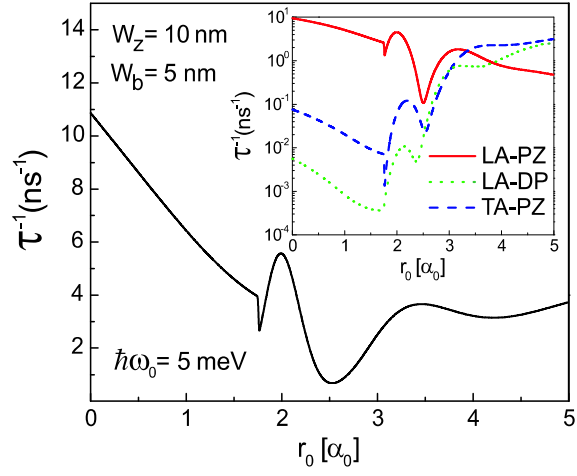}
\protect\caption{(Color online) (a) The total relaxation rate for
the transition from the first excited state to the ground state as a
function of the ring radius; (b) the LA-DP, LA-PZ and TA-PZ
contributions.} \label{fig4}
\end{center}
\end{figure}

In Fig. \ref{fig4} we study the electron relaxation in CQRs in the
case $B=0$ as a function of the ring radius $r_0$ at $B=\,$0. In
particular, we deal with the CQRs with $\hbar \omega_0=5$ meV and
$W_z=10$ nm. This choice is justified by the fact that in
experimental realization of QRs the radius is found to be $r_0 \sim
15 \div 60\,$ nm and the ring thickness is found to be $W_z \sim
10\,$nm \cite{lorke,emperador,mano}. In the rest of the paper we
will consider rings with such values of lateral and vertical
confinements. The total relaxation time for the transition from the
first excited to the ground state is shown in the picture, while the
contributions arising from the different scattering mechanisms are
shown in the inset. All the curves present a discontinuity
corresponding to the $(0,0,1)$ and $(0,\pm1,0)$ crossings. The total
relaxation time is of the order of fractions of nanoseconds and presents a
modulation with varying radius. More interesting information
comes from observing the different contributions from DP and PZ
couplings: for small ring radius the LA-DP scattering is order of
magnitude larger than the LA-PZ and TA-PZ scatterings, in agreement
with the case of CQDs. In this case the PZ coupling can be
disregarded: this is the reason why DP coupling is often the only
source of decoherence considered in the literature. On the other
hand, PZ scattering is much larger than DP for large $r_0$. There is
a clear crossover from which the PZ coupling starts to dominate over
the DP one ($r_0 > 3 \alpha_0$). This findings are similar to the
ones in Ref. \cite{piacente} for single QRs, so we can conclude that
the efficacy of the PZ scattering is intimately related to the ring
topology and, hence, in QRs and CQRs PZ effects are always important
and cannot be disregarded.

\begin{figure}[h]
\begin{center}
\includegraphics [width=7.5cm]{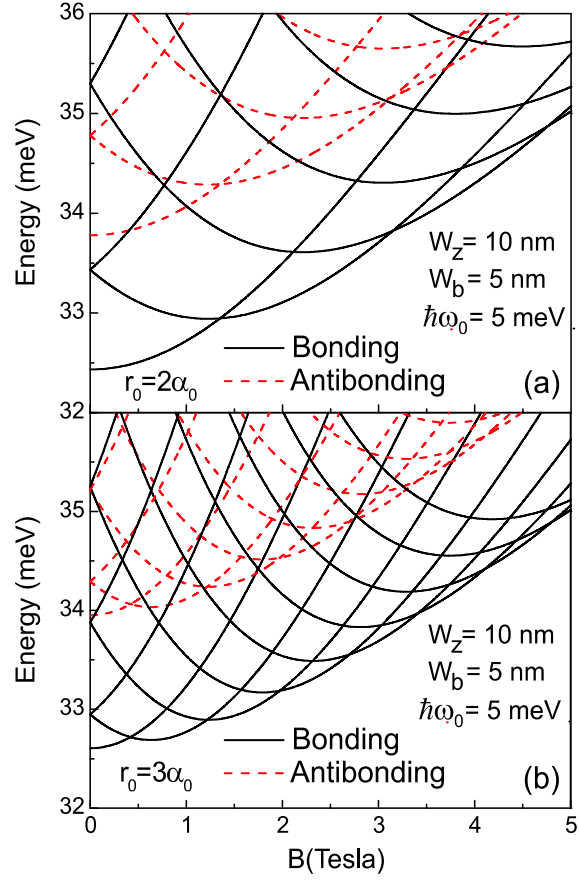}
\caption{(Color online) The energy spectrum as a function of the
external magnetic field: (a) for $r_0=2\alpha_0$ and (b)
$r_0=3\alpha_0$. The solid lines represent bonding states, while the
dashed lines are relative to antibonding states.} \label{fig5}
\end{center}
\end{figure}

When a vertical magnetic field is present new effects appear. First
of all, as one can see in Fig. \ref{fig5}(a) and (b) the typical
level crossings of multiple connected geometries appear in the
energy spectra. The ground state, for instance, changes from the
state with $m=0$ to the ones with $m=-1,-2,-3,...$. It is worth stressing that this feature is not present in QDs and CQDs, where no
level crossings for the ground state occur, even in the presence of
strong magnetic fields, and the low-lying levels converge to the
first Landau levels without crossings. Actually, identifying the
crossing points in the energy spectra gives straightforward
information about the specific ring topology
\cite{JClimentePRB,fuhrer}. In the presence of magnetic field the
number of crossing points is strongly increasing with increasing
$r_0$, as is evident in comparing Figs. \ref{fig5}(a) and \ref{fig5}(b). Moreover, the low-lying levels are bonding states
when a vertical magnetic field is applied.

\begin{figure}[h]
\begin{center}
\includegraphics [width=7.5cm]{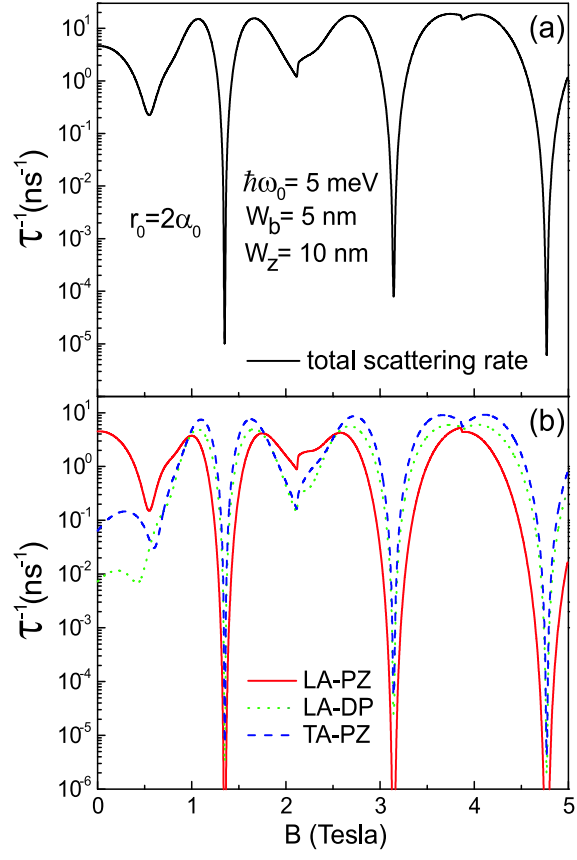}
\protect\caption{(Color online) (a) The total relaxation rate from
the first excited state to the ground state as a function of the
magnetic field for a ring of radius $r_0=2\alpha_0$; (b) the LA-DP,
LA-PZ and TA-PZ contributions.} \label{fig6}
\end{center}
\end{figure}

\begin{figure}[h]
\begin{center}
\includegraphics [width=7.5cm]{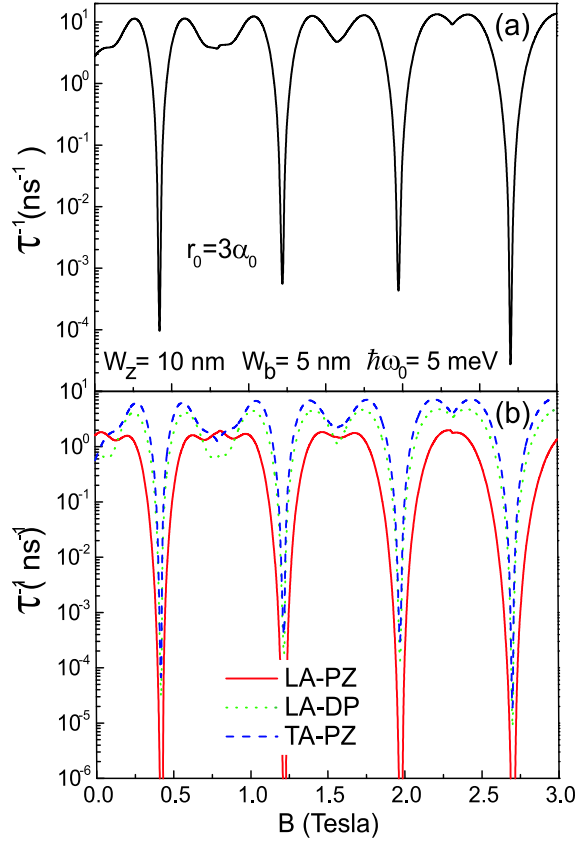}
\protect\caption{(Color online) The same as Fig. \ref{fig6}, but
with $r_0=3\alpha_0$.} \label{fig7}
\end{center}
\end{figure}

For fixed ring radii $r_0=2\alpha_0$ and $r_0=3\alpha_0$, we
calculate the scattering rates among the first excited state to the
ground state as a function of magnetic field. The results are shown
in Figs. \ref{fig6}(a) and \ref{fig7}(a), respectively. The
contributions of the DP and PZ phonons to the scattering rates are
given separately in Figs. \ref{fig6}(b) and \ref{fig7}(b). The
first notable features is that in both cases the total relaxation
rate shows oscillations with striking dips corresponding to the
level crossing points. This property is related to the fact that the
scattering matrix elements $| \langle \phi_{n',m'}|
M^2_{\lambda}(\overrightarrow {\bf q }) | \phi_{n,m}\rangle |^2$
vanish in correspondence of a crossing point, where $q_0=0$. The
number of dips increases with increasing radius, because the number
of level crossings is larger for larger radius. Furthermore, the PZ
and DP phonon scattering rates have approximatively the same
magnitude and they both contribute to the total scattering rate.
This is very different from the CQD case. Actually, in CQDs in the
presence of vertical magnetic fields up to 5 Tesla the PZ rates are
always orders of magnitude smaller than the DP ones
\cite{ClimentePRB}. Once more, this confirms and stresses that the
scattering mechanism is significantly different between rings and
dots and that such a difference is amplified when a magnetic field
is applied. In order to describe properly the phonon scattering
processes in the CQRs in magnetic fields, inclusion of the PZ
interaction is, therefore, fundamental.

\begin{figure}[h]
\begin{center}
\includegraphics [width=9cm]{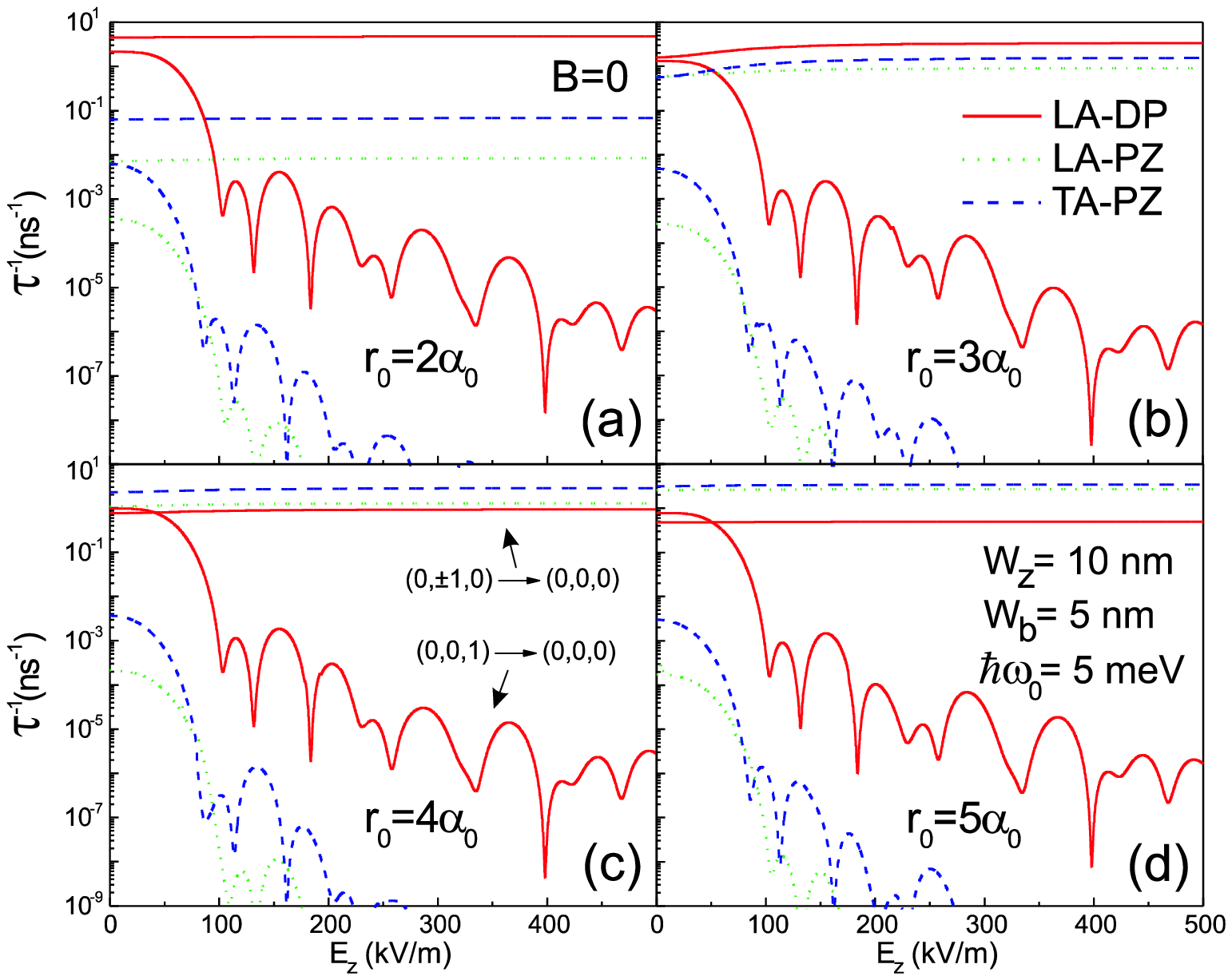}
\protect\caption{(Color online) The different scattering mechanisms
as a function of vertical electric field for the
$(0,0,1)\rightarrow(0,0,0)$ and $(0,\pm 1,0)\rightarrow(0,0,0)$
transitions for different ring radii.} \label{fig8}
\end{center}
\end{figure}

Finally, we study the effect of an external electric field along the
vertical direction. In Fig. \ref{fig8} we depict the $E_z$-dependent
scattering rate in CQRs with different ring radii in the absence of
magnetic field. Applying a vertical electric field is a efficient
way to tune the tunneling, therefore it influences the transition
between states with different parities. In fact, we find strong
oscillations of the DP and PZ scattering rates for the
$(0,0,1)\rightarrow(0,0,0)$ transition, while the scattering rates
for the $(0,\pm1,0)\rightarrow(0,0,0)$ transition are almost
constant. Interestingly, the PZ scattering rate decays much faster
than that of the DP in the $(0,0,1)\rightarrow(0,0,0)$ transition.
Although PZ and DP scatterings have about the same magnitude in the
absence of external fields for relatively small ring radii, the
application of an electric field soon turns DP into the dominant
relaxation mechanism and drastically reduces the PZ scattering, as
one can see in Figs. \ref{fig8}(a) and \ref{fig8}(b). For relatively large ring
radii [see Figs. \ref{fig8}(c) and \ref{fig8}(d)] the situation is different and
more complex. As a matter of fact for the
$(0,\pm1,0)\rightarrow(0,0,0)$ transition the PZ scattering is
larger than the DP one, while for $(0,0,1)\rightarrow(0,0,0)$
transition the DP relaxation rate is order of magnitude larger than
the the DP one.

\section{Conclusions}\label{sec4}

We studied the acoustic phonon induced electron relaxation in vertically
coupled GaAs quantum rings, which is of fundamental importance for
applications of quantum rings as quantum gates and nanodevices. We
investigated how ring geometry, inter-ring tunneling, and external
fields affect the electron-phonon scattering and the relaxation from
the first excited state to the ground state. Our calculations show
that the electron-phonon scattering strongly depends on the ring size,
separation between the two QRs, and on external fields. We took into
account both deformation potential and piezoelectric field couplings
and demonstrated that they both give important contributions to the
electron relaxation. Piezoelectric interactions, often neglected in
the literature in the studies of relaxation in quantum dots and
coupled quantum dots, represent the major source of scattering in
the presence of external magnetic field and for large ring radius
and/or large ring separation, while for small ring radius and zero
external field deformation potential couplings prevail. Furthermore,
we have shown that significant oscillations in the scattering rates
from electron excited states to the ground state are present and
depend on either the geometry of the structure or the external
fields.

\section{Acknowledgments}

This work was supported by FAPESP and CNPq, Brazil

\newpage

\begin{thebibliography}{*}

\bibitem{garcia} J.~M. Garcia, G. Medeiros-Ribeiro, K. Schmidt, T.
Ngo, J.~L. Feng, A. Lorke, J.~P. Kotthaus, and P.~M. Petroff, Appl.
Phys. Lett. {\bf71}, 2014 (1997).

\bibitem{lorke} A. Lorke, R.~J. Luyken, A.~O. Govorov, J.~P.
Kotthaus, J.~M. Garcia, and P.~M. Petroff, Phys. Rev. Lett. {\bf84},
2223 (2000).

\bibitem{mailly} D. Mailly, C. Chapelier, and A. Benoit, Phys. Rev.
Lett. {\bf70}, 2020 (1993).

\bibitem{mano} T. Mano, T. Kuroda, S. Sanguinetti, T. Ochiai, T.
Tateno, J. Kim, T. Noda, M. Kawabe, K. Sakoda, G. Kido, and N.
Koguchi, Nano Lett. {\bf5}, 425 (2005).


\bibitem{Planelles} J. Planelles, F. Rajadell, J.~I. Climente, M.
Royo, and J.~L. Movilla, J. Phys.: Condens. Matter {\bf 17}, 1573
(2005).

\bibitem{ManoPRB} T. Kuroda, T. Mano, T. Ochiai, S. Sanguinetti, K.
Sakoda, G. Kido, and N. Koguchi, Phys. Rev. B {\bf 72}, 205301
(2005).

\bibitem{SzafranPRB} B. Szafran and F.~M. Peeters, Phys. Rev. B {\bf
72} 155316 (2005).

\bibitem{PlanellesEPJB} J. Planelles and J.~I. Climente, Eur. Phys. J.
B {\bf 48}, 65 (2005).

\bibitem{GranadosAPL} D. Granados, J.~M. Garcia, T. Ben, and
S.~I.Molina, Appl. Phys. Lett. {\bf 86}, 071918 (2005).

\bibitem{JClimentePRB} J.~I. Climente and J. Planelles, Phys. Rev. B
{\bf 72}, 155322 (2005).

\bibitem{MaletPRB} F. Malet, M. Barranco, E. Lipparini, R. Mayol, M.
Pi, J.~I. Climente, and J. Planelles, Phys. Rev. B {\bf 73}, 245324
(2006).

\bibitem{Castelano} L.~K. Castelano, G.~Q. Hai, B. Partoens, and
F.~M. Peeters, Phys. Rev. B {\bf 74}, 045313 (2006).

\bibitem{hun} K.~H. Ahn and P.
Fulde, Phys. Rev. B {\bf 62}, R4813 (2000).

\bibitem{bayerSC} M. Bayer, P. Hawrylak, K. Hinzer, S. Fafard, M.
Korkusinski, Z. R. Wasilewski, O. Stern, and A. Forchel, Science
{\bf 291}, 451 (2001).

\bibitem{burkard} G. Burkard, G. Seelig, and D. Loss, Phys. Rev. B
{\bf 62}, 2581 (2000).

\bibitem{BastardPRB} U. Bockelmann and G. Bastard, Phys. Rev. B {\bf
42}, 8947 (1990).

\bibitem{FujisawaNAT} T. Fujisawa, D.~G. Austing, Y. Tokura, Y.
Hirayama, and S. Tarucha, Nature (London) {\bf 419}, 278 (2002).

\bibitem{OrtnerPRB} G. Ortner, R. Oulton, H. Kurtze, M. Schwab, D.~R.
Yakovlev, M. Bayer, S. Fafard, Z. Wasilewski, and P. Hawrylak, Phys.
Rev. B {\bf 72}, 165353 (2005).

\bibitem{ClimentePRB} J.~I. Climente, A. Bertoni, G. Goldoni, and E.
Molinari, Phys. Rev. B {\bf 74}, 035313 (2006).

\bibitem{piacente} G. Piacente and G.~Q. Hai, Phys. Rev. B {\bf 75}, 125324
(2007).

\bibitem{granadosapl} D. Granados, J.~M. García, T. Ben, and S.~I. Molina,
Appl. Phys. Lett. {\bf 86}, 071918 (2005).

\bibitem{suareznano} F. Su\'{a}rez, D. Granados, M.~L. Dotor and J.~M. García,
Nanotechnology {\bf 15}, S126 (2004).

\bibitem{simoninPRB} J. Simonin, C.~R. Proetto, Z. Bartevica, and G.
Fuster, Phys. Rev. B {\bf 702}, 205305 (2004).

\bibitem{HaiAPL} G.~Q. Hai and S.~S. Oliveira, Appl. Phys. Lett.
{\bf88}, 196101 (2006).

\bibitem{Mahan} G.~D. Mahan, {\em Many-Particle Physics} (Plenum
Press, New York, 1990).

\bibitem{blackmore} J.~S. Blakemore, J. Appl. Phys. {\bf 53}, R123
(1982).

\bibitem{zanardiprl} P. Zanardi and F. Rossi, Phys. Rev. Lett. {\bf 21}, 4752 (1998).

\bibitem{emperador} A. Emperador, M. Pi, M. Barranco, and A. Lorke,
Phys. Rev. B {\bf 62}, 4573 (2000).

\bibitem{fuhrer} A. Fuhrer, S. L\"{u}scher, T. Ihn, T. Heinzel, K.
Ensslin, W. Wegschel, and M. Bichler, Nature (London){\bf 413}, 822 (2001).

\end{thebibliography}
\end{document}